\documentstyle[12pt,twoside,fleqn,espcrc1]{article}
\def\lsim{\raise0.3ex\hbox{$<$\kern-0.75em\raise-1.1ex\hbox{$\sim$}}}
\def\gsim{\raise0.3ex\hbox{$>$\kern-0.75em\raise-1.1ex\hbox{$\sim$}}}
\newdimen\digitwidth\setbox0=\hbox{\rm0}\digitwidth=\wd0
\def\S{{\scriptscriptstyle S}} \def\T{{\scriptscriptstyle T}}
\def\ALPS{\alpha_\S}\def\alps{\ifmmode\ALPS\else$\ALPS$\fi}
\def\PT{p_\T}\def\pt{\ifmmode\PT\else$\PT$\fi}
\def\PPB{\langle\overline\psi\psi\rangle}\def\ppb{\ifmmode\PPB\else$\PPB$\fi}

\def\etal{{\sl et al.\/}}

 \title{Electromagnetic Signals and Backgrounds in Heavy-Ion Collisions}

 \author{Sourendu Gupta
         \address{HLRZ, c/o KFA J\"ulich, D-5170 J\"ulich, Germany}}

\begin{document}
\hfill HLRZ 53/93\ \ \ \ \ \ \\
{\vskip2cm\center
  {\large{Electromagnetic Signals and Backgrounds}\par}
  {\large{in Heavy-Ion Collisions}\par}
  \vskip1cm{\normalsize{Sourendu Gupta}\par}
  {HLRZ, c/o KFA J\"ulich, D-52425 J\"ulich, Germany}\par\vskip3cm}
 \begin{abstract}
Aspects of the dilepton spectrum in heavy-ion collisions are discussed, with
special emphasis on using lattice computations to guide the phenomenology
of finite temperature hadronic matter. The background rates for continuum
dileptons expected in forthcoming experiments are summarised. Properly
augmented by data from ongoing measurements at HERA, these rates will serve
as a calibrating background for QGP searches. Recent results on the temperature
dependence of the hadronic spectrum obtained in lattice computations below the
deconfinement transition are summarised. Light vector meson masses are strongly
temperature dependent. Accurate measurements of a resolved $\rho$-peak in
dimuon spectra in present experiments are thus of fundamental importance.
 \end{abstract}
\vfil
{\sl Invited talk at the Quark Matter '93 Conference, Borl\"ange, Sweden,
     June 1993}
 \maketitle

 \begin{abstract}
Aspects of the dilepton spectrum in heavy-ion collisions are discussed, with
special emphasis on using lattice computations to guide the phenomenology
of finite temperature hadronic matter. The background rates for continuum
dileptons expected in forthcoming experiments are summarised. Properly
augmented by data from ongoing measurements at HERA, these rates will serve
as a calibrating background for QGP searches. Recent results on the temperature
dependence of the hadronic spectrum obtained in lattice computations below the
deconfinement transition are summarised. Light vector meson masses are strongly
temperature dependent. Accurate measurements of a resolved $\rho$-peak in
dimuon spectra in present experiments are thus of fundamental importance.
 \end{abstract}

\section{Introduction}

Electromagnetic probes of the quark gluon plasma have been surveyed
extensively in the last few years. I will not repeat the material
covered by these excellent reviews \cite{revs}. Since there has not been much
development in the theory of photon signals since the last Quark Matter
meeting, therefore, in the rest of this talk I will concentrate
on dilepton signals and backgrounds.

Recall that mass spectra for opposite sign dileptons form
a continuum with conspicuous resonances sitting over it. The cross
section is very closely related to a theorist's favourite quantity--- the
spectral density of a vector correlation function. All observed resonances
correspond to flavour singlet vector mesons. With sufficient mass resolution
in the spectra the fate of each such meson can be seen in the dense and,
possibly, thermalised hot matter formed as a result of heavy ion collisions.
The ease with which individual resonances can be isolated and
studied by well-designed experiments makes the dilepton signal
a tool which is neglected only by the most foolhardy physicist.
The continuum itself may be interesting for various reasons, many of which
have been reviewed before.

I shall spend most of my allotted pages on scenarios which are built for
matter in, or not far from, thermal equilibrium. The main reason for this
emphasis is the ease with which theorists can do these computations; but,
as reported in this meeting \cite{geig,gyu}, there are model computations
now which indicate a fairly short thermalisation times in heavy-ion
collisions. Nevertheless, it is necessary to keep in mind that the dense
systems formed may spend a significant fraction of their lifetimes trying
to come to a state of equilibrium. If they succeed, they will be doing
much better than most people.

Uptil now very little work has been done with non-equilibrium scenarios. It
should be mentioned that Shuryak's two-step thermalisation model \cite{shur}
is an attempt at constructing a toy model of non-equilibrium phenomena.
Other such attempts are hydrodynamical shock waves and burning walls
\cite{kaja}, swiss-cheese instabilities \cite{svet}, {\sl etc\/}.
All these dynamical modes can be married to standard dilepton production
processes, thereby yielding possible signals of non-equlibrium phenomena.
All these modes yield continuum dileptons. However it is known that
non-equilibrium dynamics can give rise to narrow peaks in spectral densities
which have no relation to the equilibrium energy levels of the system
\cite{hyst}. This feature seems so generic that one wonders whether such a
peak may not be seen in heavy-ion collisions. The moral I want to draw is
that experiments should keep watch for phenomena which theories cannot yet
deal with. Non-equilibrium statistical physics is a growing new branch
of fundamental physics and heavy-ion experiments can make important
contributions here.

To turn to concrete physics, I shall divide my talk into two major parts.
The first will be concerned with continuum dileptons in various mass regions;
the second with a few selected resonances. These are discussed in the next
two sections. I shall emphasise the utility of lattice computations to guide
phenomenology. The point is that the lattice is a non-perturbative tool to
compute matrix elements in QCD which are usually obtained in models or
perturbation theory. Such work started a year ago \cite{cern} and is being
pursued further \cite{hlrz}.

\section{The Continuum}
Continuum dileptons may be useful as a probe for thermal matter. All
computations of the thermal signal to date have been performed in high
temperature perturbation theory. Lattice computations give clear evidence
for non-perturbative phenomena at temperatures close to $T_c$. It would,
therefore, be interesting to obtain non-perturbative lattice estimates
for the relevant QCD matrix elements at high temperature. This is now
being done; the results will be available in the near future.

I shall speak of the high mass ($M\gsim5$ GeV), low mass ($M\lsim1$ GeV) and
the intermediate mass region. Rougly speaking, the intermediate mass region
is bounded by the $\rho$ and $J/\psi$ resonances. I shall also speak of the
region with $M\gsim10$ GeV as the very-high mass region. This is the region
beyond the $\Upsilon$ resonances.

\subsection{Low mass continuum}
In hadron-hadron collisions, the low-mass continuum seems to be understood
in terms of several processes--- Dalitz pairs, bremsstrahlung from pions
{\sl etc\/}. There have been quantitative attempts to extend this picture to
heavy-ion collisions \cite{satz}. The low-mass dilepton spectrum may be
strongly dependent on properties of thermalised hadronic matter, specially
if the thermalisation time turns out to be close to this year's favourite
number--- a fraction of a fermi. A first attempt at such estimates now
exists \cite{turko}. Such computations need, quite crucially, input on
hadronic masses, widths and interaction strengths at finite temperatures.
These can be obtained in lattice computations, and I shall summarise a
recent computation \cite{hlrz} in the next section.

\subsection{High mass continuum}
Even if thermalisation times are short, the high mass continuum cross section
consists of pre-equilibrium processes. The very high mass region is expected
to consist essentially of Drell-Yan pairs at both the LHC and RHIC energies.
At LHC, in the mass range between the $\Upsilon$ and $J/\psi$, there could
be substantial contribution from open bottom production. This has yet to be
estimated.

A state-of-the-art computation of rates for LHC and RHIC was presented in the
Aachen Workshop in 1990 \cite{aachen}. These were based on exponentiated
${\cal O}(\alps)$ cross sections computed in perturbative QCD \cite{dyalps}.
Needless to say, these cross sections agree extremely well with data obtained
in the range of energies $19.4\le\sqrt S\le640$ GeV \cite{expt}. There have
been two advances since the last estimates were made. One is theoretical---
the full ${\cal O}(\alps^2)$ Drell-Yan cross section has now been computed
\cite{dyexp2}. In the high mass region these do not affect the old estimates.
The second advance is experimental--- measurements of structure functions in
the range $x\le10^{-3}$ have now been performed \cite{h1}. These have
possible effect on estimates of cross sections at LHC energies; and must be
taken into account once the data from HERA has been analysed.

The main kinematic features of the Drell-Yan cross section are the following.
With increasing $\sqrt S$, the rapidity distribution of Drell-Yan pairs
develops a plateau. At RHIC energies this is 3 units wide; at LHC energies
the width increases to about 5 units. With increasing $\sqrt S$, at fixed $M$,
the increase in the \pt-integrated cross section comes from the growth
of the perturbative tail in the \pt-distribution. A consequence of this is
a roughly linear growth of $\langle\pt^2\rangle$ with $S$ which occurs
for $\sqrt S\gsim50$--100 GeV. Detailed predictions are given in \cite{aachen}.

It should be remembered that there are no mass scales involved in the QCD
predictions apart from $\Lambda_{QCD}$. Thus absolute normalisations of
cross sections, $\langle\pt^2\rangle$, {\sl etc\/}, and their dependence on
$\sqrt S$ are predictions. Approximate Monte-Carlo schemes \cite{geig},
on the other hand, contain various mass parameters in momentum
cutoffs. Thus absolute normalisations and other such dimensional quantities
are fitted at each $\sqrt S$ seperately. As a result of this, it is necessary
that these Monte Carlo generators be compared with a proper Drell-Yan
estimate and data at all relevant values of $\sqrt S$.

\subsection{Intermediate mass continuum}
The intermediate region of the dilepton continuum comes from a complex mixture
of sources. This is probably the most poorly understood part of the continuum
spectrum. It is also the most important region for the continuum thermal
dilepton signal. If the initial temperature is around 250 MeV, then it turns
out that the thermal signal vanishes below the extrapolated Drell-Yan cross
section at $M\approx2$--$2.5$ GeV. If the initial temperature turns out to be
about 1 GeV, then this signal may be visible above the same background even
for $M\approx5$ GeV \cite{larry}.

The extrapolation of Drell-Yan cross sections into this region suffers from
two main ambiguities. The strongest source of uncertainty is in the parton
luminosities at small $x$. At LHC, the range of masses of interest corresponds
to $x\approx5\times10^{-4}$. New physics may come into play in this region of
kinematics. Structure function measurements now being done at HERA will be
crucial for a better understanding of this region. A second uncertainty is
in the importance of higher order resummed perturbative corrections. The new
results mentioned above show that these are under control.
I estimate an uncertainty
of a factor of three when extrapolating Drell-Yan results to the intermediate
mass range at RHIC and LHC energies. The thermal signal turns out to have a
much steeper slope than the Drell-Yan continuum. Hence this large uncertainty
still yields a small error (less than 0.5 GeV) in the cutoff mass below
which the thermal signal dominates over this background.

It should be remembered that the Drell-Yan process is only one of the many
backgrounds in the dilepton channel in the intermediate mass region. At the
higher end of this region decays of heavy-flavour quarks give rise to a large
dilepton rate. Such a contribution had been pointed out by Shor long back.
A minijet computation \cite{physlet} for the process
\begin{equation}
A + B \rightarrow {\rm jets} \rightarrow c (b) \rightarrow {\rm leptons},
\label{minijet}
\end{equation}
shows that one should expect a large number of single leptons per event. These
combine into a large dilepton background. This is relatively innocuous, since
the rate for unlike sign is the same as for like sign pairs, and thus can be
subtracted. A detailed Monte Carlo study is reportedly in progress
\cite{hijing}. More problematic is the background from the cascade decay of
bottom into strange with unlike sign pairs. This background also needs to be
computed.

At the lower end of the intermediate mass range the situation is even more
complicated. The processes which contribute in heavy-ion collisions have
probably not been completely enumerated yet. NA36 has some new data which
they will discuss in this meeting \cite{na36}.

\subsection{The deconfined phase}
The continuum dimuon cross section in the deconfined phase is the signal for
which the processes discussed in the previous subsections are the background.
It is customary to compute this cross section in high-temperature perturbation
theory. The lattice can furnish cross checks on this procedure.

In recent years several studies of lattice QCD \cite{old,cern} have
furnished evidence that the high temperature phase really consists of
deconfined quarks. Thus the primary condition for perturbation theory seems
to be validated--- the degrees of freedom are correctly identified.

However, there are indications that these quarks, under certain circumstances,
have fairly strong self-interactions. A study \cite{cern} has clarified this
situation. For mass scales below the Debye screening mass, one could write an
effective theory for the quarks in the form
\begin{equation}
S_{eff}\;=\; \int d^4x \left[ \overline\psi(\gamma_\mu\partial_\mu+m)\psi
    +\sum_\Gamma g_\Gamma \left(\overline\psi\Gamma\psi\right)^2
    +\cdots\right],
\label{effective}
\end{equation}
where $\Gamma$ denotes a direct product of spin and flavour matrices, and the
sum is over the whole set of such products. The ellipsis denote neglected terms
of higher mass dimension. From lattice measurements it has been found
\cite{cern} that, although the effective couplings $g$ in the scalar and
pseudo-scalar channels are large, those in the vector and pseudo-vector
channels are rather small already at temperatures close to $T_c$.
Thus, this observation implies that perturbative computations of
dilepton and photon production rates may be reliable quite close to the phase
transition temperature. A similiar computation in unquenched QCD is now in
progress.

For $T<1.2T_c$, however, perturbation theory does
break down. This is reflected in the growth of all the effective couplings as
one approaches $T_c$ from above. It may be possible to use lattice measurements
to obtain the matrix element relevant to photon or dilepton cross sections.
Such studies are planned.

\section{The Resonances}

Heavy-quark resonances have been the subject of concerted study for the
last five years. The situation is slowly being clarified; there is new
and exciting data this year from the NA38 collaboration \cite{na38}.
Lighter resonances have been studied in models for many years now. There
is exciting news on these from recent lattice computations.

\subsection{Charmonium}
Based on lattice studies of the static inter-quark potential, heavy-quarkonia
have been suggested as a signal for screening. Screening sets in at the QCD
phase transition and the screening length decreases with increasing $T$.
Consequently, different resonances are suppressed to different extent under
the same physical circumstances. This year's result from NA38 \cite{na38}
shows a strong $E_\T$-dependence to the relative suppression between the
$\psi'$ and the $J/\psi$. The data is compatible with estimates given in
\cite{psi} as well as in \cite{sean}.

\begin{figure}[htb]\vskip42mm

\begin{minipage}[t]{80mm}
\includegraphics{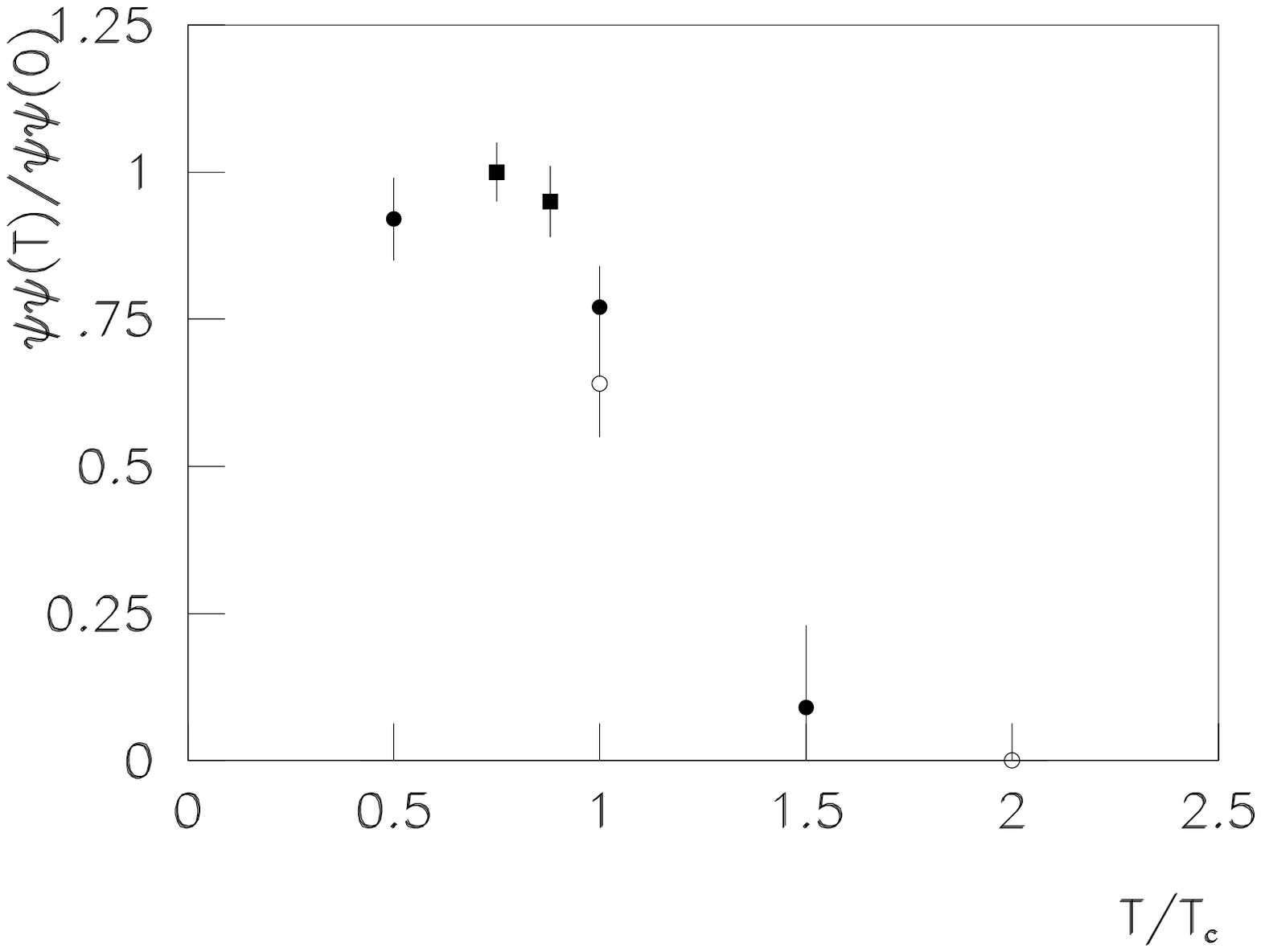}
\label{fig:ppb}
\caption{The temperature dependence of \ppb, for quenched simulations with
         $N_\tau=4$ (filled circles) and 8 (squares) and from a 4-flavour
         simulation with $N_\tau=8$ (open circles).}
\end{minipage}
\hspace{\fill}
\begin{minipage}[t]{75mm}
\includegraphics{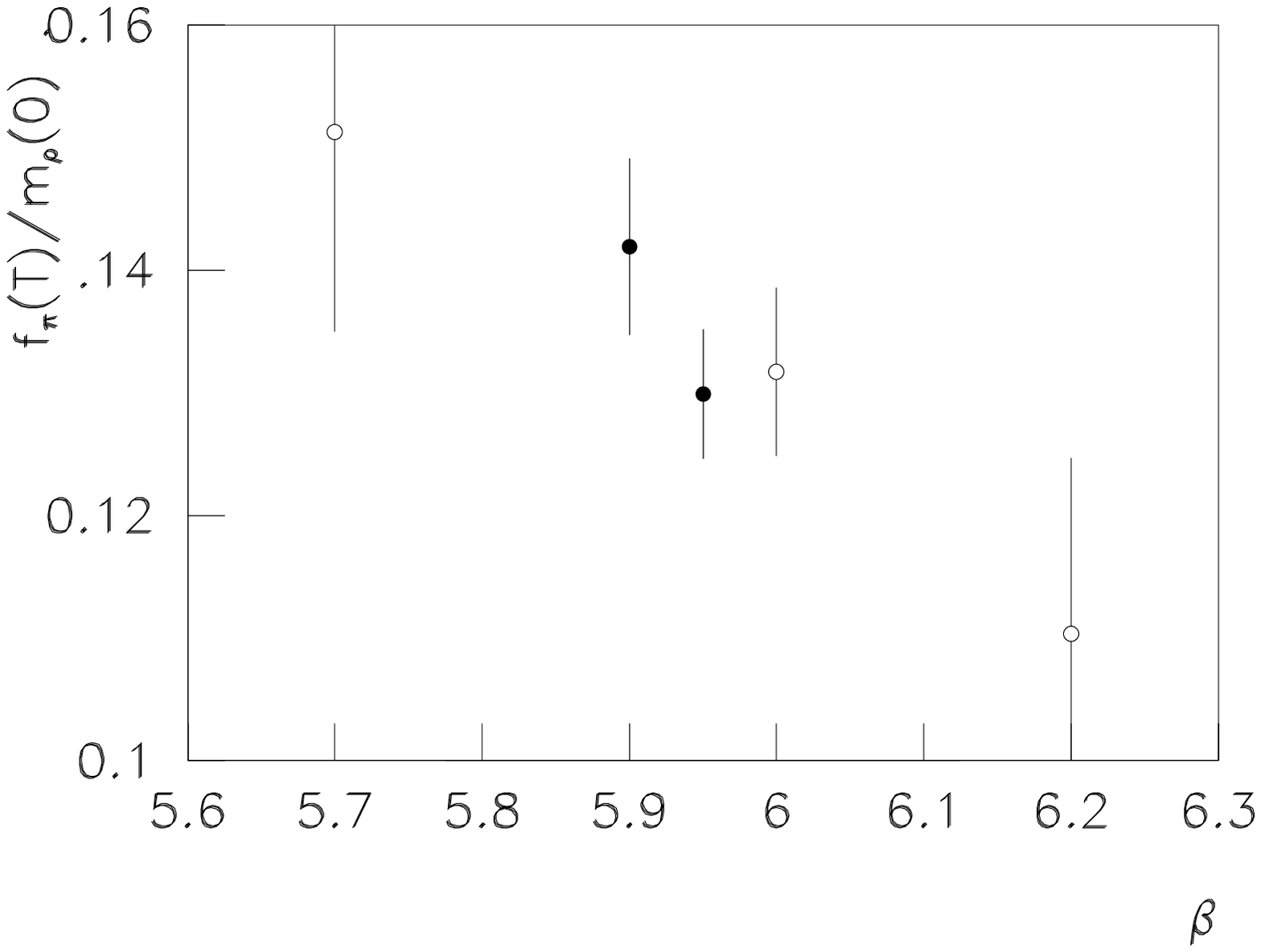}
\label{fig:fpi}
\caption{The temperature dependence of $f_\pi$. Data for $T=0$ (open circles)
         and at finite temperatures (filled circles) $T\approx0.75T_c$
         ($\beta=5.9$) and $T\approx0.9T_c$ ($\beta=5.95$).}
\end{minipage}

\end{figure}

\subsection{The $\rho$ meson}
Two recent studies of quenched lattice QCD have concentrated on hadronic
properties for $0<T<T_c$. One of these \cite{cern} was done on $N_\tau=4$
lattices on very large spatial volumes, extending to $(8/T)^3$, at $0.5T_c$.
Work now in progress \cite{hlrz} extends these computations to $N_\tau=8$
on spatial volumes of $(4/T)^3$ at $0.75T_c$ and $0.88T_c$. In both these
studies the values of the quark condensate, \ppb, pion decay constant,
$f_\pi$, and the pion and $\rho$ masses have been studied. The temperature
dependence of these quantities is obtained by comparison with $T=0$
measurements at the same lattice spacing.

It is known that the quark condensate goes to zero with a discontinuity at
$T_c$ in both the quenched \cite{cern,hlrz} and 4-flavour \cite{mtc} theories.
In Figure 1 we show the measured temperature dependence of
$\ppb(T)/\ppb(0)$ (the $T=0$ values are taken from \cite{salinas}).
Two features bear comment. First, note that the discontinuity at $T_c$ is
similiar in the two cases. Second, \ppb{} seems to be relatively temperature
independent up to $T\approx0.9T_c$.

A non-vanishing quark condensate implies a vanishing pion mass in the chiral
limit. The physical pion mass is obtained from the relation
\begin{equation}
m_\pi^2 \;=\; A_\pi m_q.
\label{mpi}
\end{equation}
Here $m_q$ is the quark mass. Measurements of $A_\pi$ on the lattice, thus
give information on the temperature dependence of the pion mass. Our
measurements reveal no change in $A_\pi$ compared to the values at $T=0$
for temperatures up to $0.9T_c$ (see Figure 3a). Consequent
to these two facts, the pion decay constant, $f_\pi$, shows no change with
temperature upto $T=0.9T_c$. This is explicitly shown in Figure 2,
using the data of \cite{hlrz} and the $T=0$ data of \cite{rajan}.
In Figures 1 and 2, mass ratios have been used in order to
remove most lattice effects.

\begin{figure}[htb]\vskip42mm

\begin{minipage}[t]{80mm}
\includegraphics{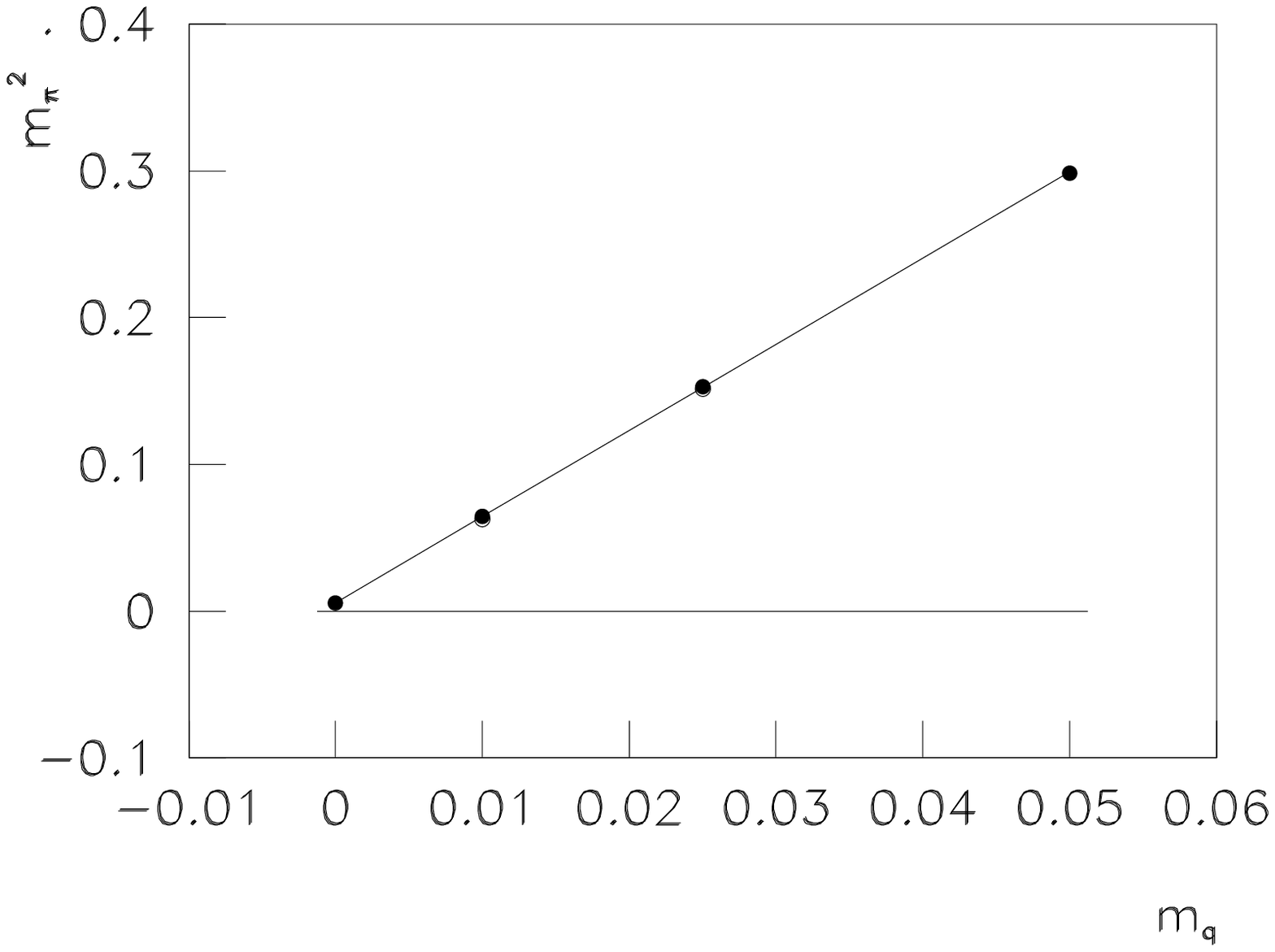}
\end{minipage}
\hspace{\fill}
\begin{minipage}[t]{75mm}
\includegraphics{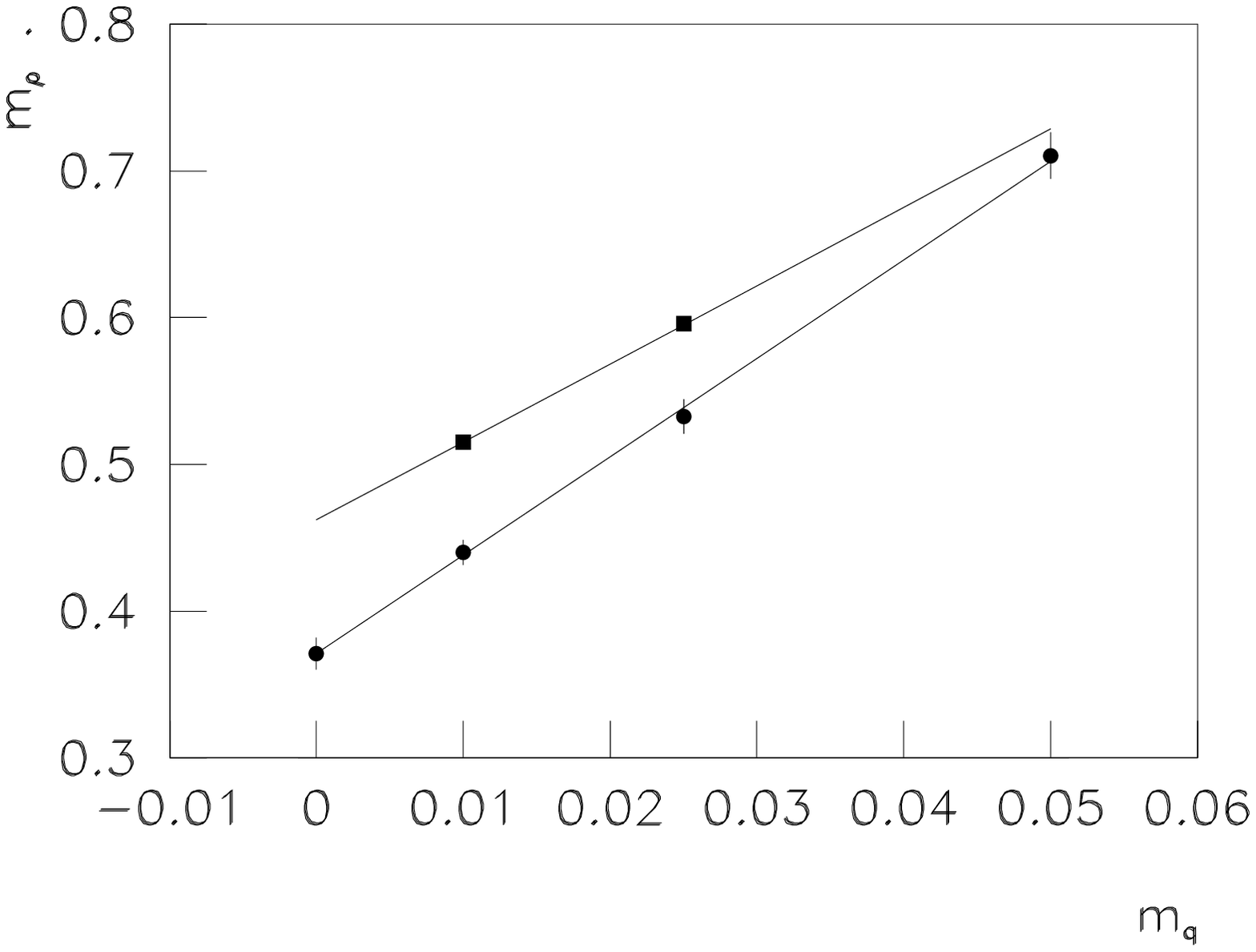}
\end{minipage}

\label{fig:mass}
\caption{The dependence of (a) $m_\pi^2$  and (b) $m_\rho$ on $m_q$ at
         $T\approx0.9T_c$ (circles) compared to data at $T=0$ (squares)
         at the same lattice spacing ($\beta=5.95$).}
\end{figure}

The value of $m_\rho$, on the other hand, seems to be quite strongly dependent
on the temperature. Measurements show that there is very little shift in the
vector meson masses at a temperature of $0.75T_c$. Within the errors of
measurement, in fact, no shift is discernible. However, at $0.9T_c$ there is
a large temperature dependent shift, visible in Figure 3b.
It is interesting that a large shift in the mass of the $\rho$ meson occurs
already at a temperature where the chiral sector of the theory sees no
temperature effect.

It should be noted that most phenomenological models of the hadron spectrum
and its temperature dependence emphasise chiral aspects of the theory. Thus
the temperature dependence of the quark condensate is one of the primary
objects of study. The vacuum of QCD, however, is characterised by many
different condensates, some invisible to the chiral sector of the theory.
In most phenomenological models, the temperature dependence of, say, the
gluon condensate is a secondary quantity, often neglected. One interpretation
of this lattice data is that some of these other condensates have strong
temperature dependence. This would imply a dynamical role for the glue
sector which is stronger than is usually assumed. Efforts to extract the
temperature dependence of at least a few of these other condensates are now
underway. If the influence of the glue sector is indeed so strong, then the
use of an universal chiral theory at finite-temperatures to obtain information
on vector and pseudo-vector mesons may not be justified.

The variation of the vector meson mass with the quark mass $m_q$ is shown in
Figure 3b at a temperature $T\approx0.9T_c$. For comparison the
corresponding data for $T=0$ at the same lattice spacing, $\beta=5.95$
\cite{gott}, is also shown. It is seen that the magnitude of the thermal
shift is dependent on $m_q$. Thus, the maximum effect is seen for the $\rho$
meson, somewhat less for the $\omega$ and $\phi$, and virtually none (at this
temperature, at least) for any heavier meson. Of course, an accurate
determination of the mass shift of a state heavier than the inverse lattice
spacing is difficult.

It is interesting to speculate what the effect of differential shifts in the
masses of the $\rho$, $\omega$ and $\phi$ mesons would be on an experiment
like NA38 which cannot resolve these seperate peaks. An obvious effect would
be to broaden this peak. Further phenomenology might be interesting.

\begin{figure}[htb]\vskip42mm

\begin{minipage}[t]{80mm}
\includegraphics{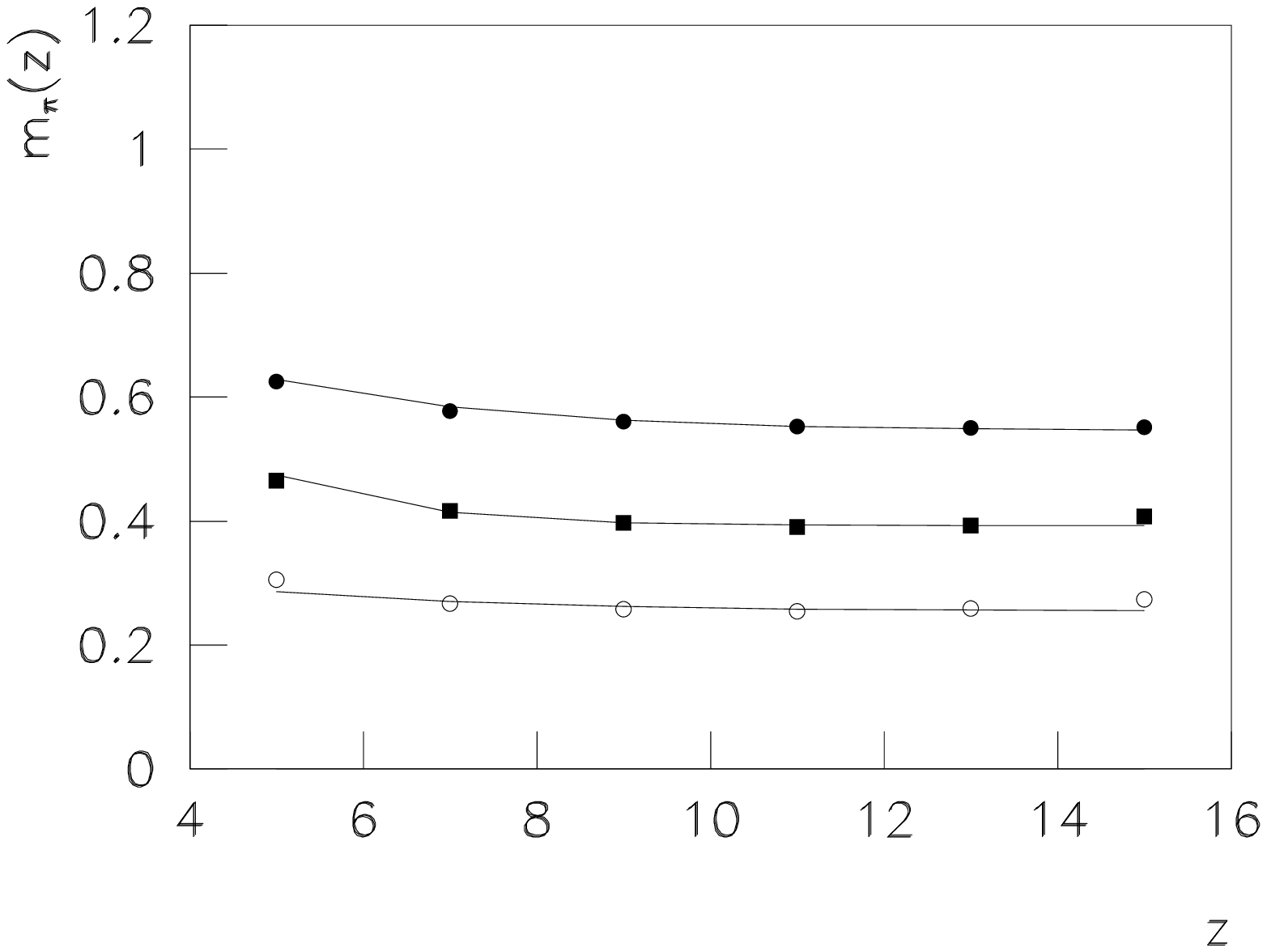}
\end{minipage}
\hspace{\fill}
\begin{minipage}[t]{75mm}
\includegraphics{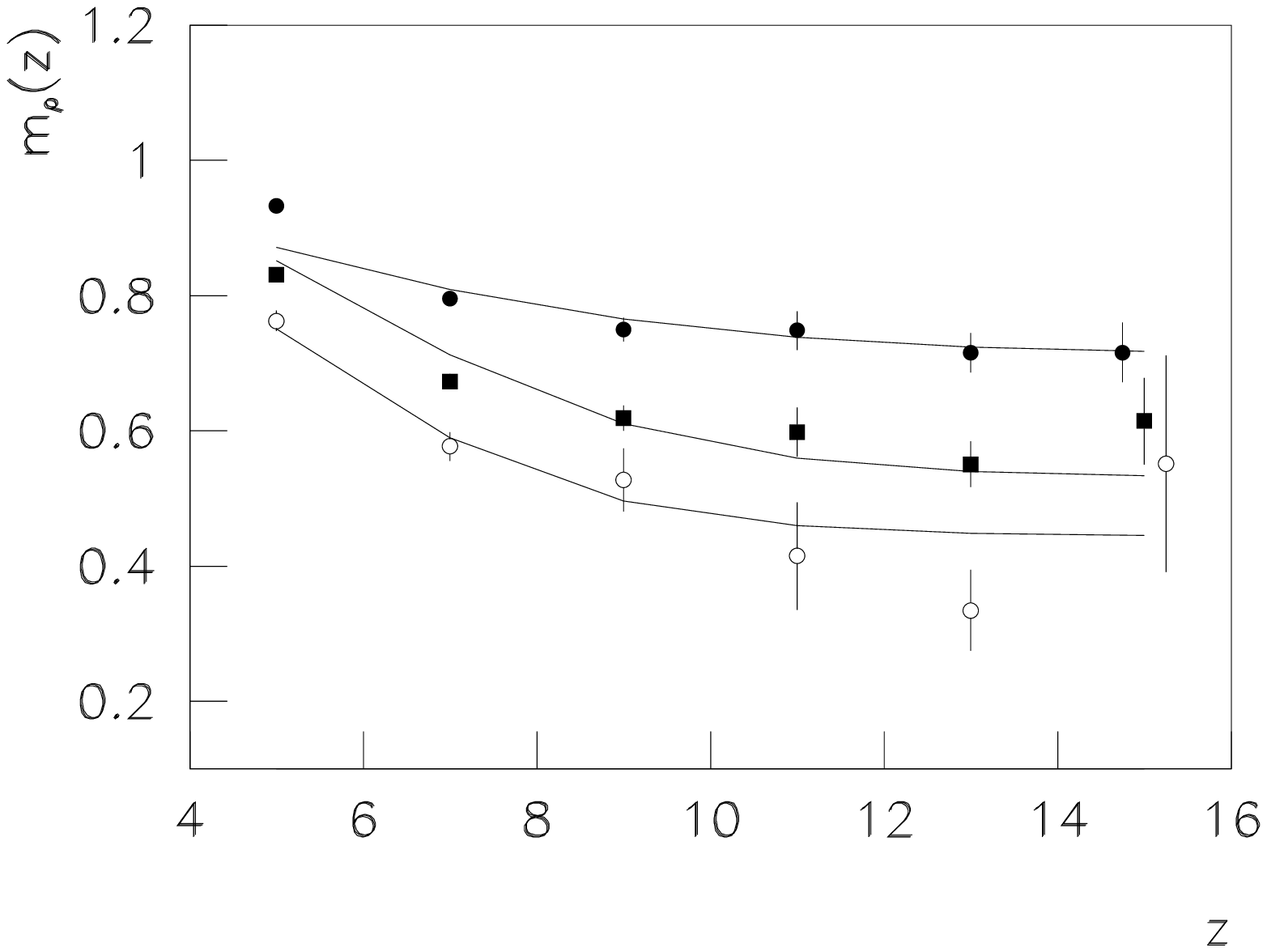}
\end{minipage}

\label{fig:localm}
\caption{Local masses for the (a) pseudoscalar and (b) vector mesons
         at $T\approx0.9T_c$ ($\beta=5.95$) for $m_q=0.05$ (filled circles),
         0.025 (squares) and 0.01 (open circles). The estimates and errors
         are obtained by jack-knife. The lines are explained in the text.}
\end{figure}

We conclude this section with some technical remarks. The masses reported here
were extracted by global fits to vector and pseudoscalar correlation functions
constructed from local sources. The well-known oscillatory behaviour in the
vector channel was suppressed by the usual stratagem of defining a correlation
function on even sites
\begin{equation}
G(2z) \;=\; {1\over2}\big[ G(2z-1)+2G(2z)+G(2z+1) \big].
\label{monotone}
\end{equation}
Local masses were extracted assuming that this correlation function can be
described by one mass, {\sl i.e.\/}, by a single hyperbolic cosine. The global
fits were made to a two-mass functional form by minimising a $\chi^2$
functional which took into account the covariance of the measurements at
different seperations. An useful cross check is to use the fitted function
to then extract `local masses' to compare with the direct measurement. Such
a comparison then checks the validity of the global fits. Example are given
in Figure 4.

{\bf Acknowledgements:} I would like to thank K.~Eskola, R.~Gavai, S.~Gavin,
A.~Irb\"ack, F.~Karsch, B.~Petersson, V.~Ruuskanen, H.~Satz, K.~Sridhar and
R.~Vogt for the discussions and/or collaborations, the results of which are
reflected in this talk.

\end{document}